\documentclass{article}
\usepackage{iclr2026_conference,times}

\iclrfinalcopy

\usepackage[colorlinks=false, pdfborder={0 0 0}]{hyperref}
\usepackage{url}
\usepackage{graphicx}
\usepackage{subcaption}
\usepackage{booktabs}
\usepackage{amsmath,amssymb,amsthm}
\usepackage[capitalize,noabbrev]{cleveref}
\usepackage{xcolor}
\usepackage{tcolorbox}
\usepackage{fancyhdr}
\usepackage{float}
\newcommand{\css}{\text{CSS}}
\newcommand{\mes}{\text{MES}}
\newcommand{\scr}{\text{SCR}}
\newcommand{\LL}{\text{LL}}

\newtcolorbox{seqbox}[1][]{colback=gray!5,colframe=gray!50,boxrule=0.5pt,left=2pt,right=2pt,top=1pt,bottom=1pt,#1}

\title{The Mechanistic Invariance Test: \\ Genomic Language Models Fail to Learn \\ Positional Regulatory Logic}

\author{Bryan Cheng$^{*}$, Jasper Zhang$^{*}$ \\
William A. Shine Great Neck South High School \\
\texttt{\{bcbc7264@gmail.com, jasperzhang1001@gmail.com\}} \\
$^{*}$Equal contribution
}

\begin{document}

\maketitle
\lhead{Workshop @ ICLR 2026}

\begin{abstract}
Genomic language models (gLMs) have transformed computational biology, achieving state-of-the-art performance in variant effect prediction, gene expression modeling, and regulatory element discovery. Yet a fundamental question threatens the foundation of this success: do these models learn the mechanistic principles governing gene regulation, or do they merely exploit statistical shortcuts? We introduce the \textbf{Mechanistic Invariance Test (MIT)}, a rigorous 650-sequence benchmark across 8 classes with scrambled controls that enables clean discrimination between compositional sensitivity and genuine positional understanding. We evaluate five gLMs spanning all major architectural paradigms (autoregressive, masked, and bidirectional state-space models) and uncover a universal failure mode. Through systematic mechanistic probing via AT titration, positional ablation, spacing perturbation, and strand orientation tests, we demonstrate that apparent compensation sensitivity is driven entirely by AT content correlation ($r$$=$0.78--0.96 across architectures), not positional regulatory logic. The failures are striking: Evo2-1B and Caduceus score regulatory elements at incorrect positions higher than correct positions, inverting biological reality. All models are strand-blind. Compositional effects dominate positional effects by 46-fold. Perhaps most revealing, a simple 100-parameter position-aware PWM achieves perfect performance (CSS$=$1.00, SCR$=$0.98), exposing that billion-parameter gLMs fail not from insufficient capacity but from fundamentally misaligned inductive biases. Larger models show stronger compositional bias, demonstrating that scale amplifies rather than corrects this limitation. These findings reveal that current gLMs capture surface statistics while missing the positional grammar essential for gene regulation, demanding architectural innovation before deployment in synthetic biology, gene therapy, and clinical variant interpretation.
\end{abstract}

\section{Introduction}

Genomic language models (gLMs) have emerged as powerful tools for understanding DNA sequence function, with applications in variant effect prediction \citep{benegas2023dna}, gene expression modeling \citep{avsec2021effective}, and regulatory element discovery \citep{ji2021dnabert}. These models---spanning transformers \citep{ji2021dnabert,dalla2023nucleotide,zhou2023dnabert2} to state-space models \citep{nguyen2023hyenadna,schiff2024caduceus}---achieve impressive predictive performance. However, a fundamental question remains: do gLMs learn \textbf{mechanistic principles} or merely memorize \textbf{statistical correlations}? We demonstrate the latter. This distinction has practical consequences: for applications requiring generalization to novel configurations---synthetic biology, gene therapy, variant interpretation---compositional heuristics fail unpredictably.

To probe this distinction, we leverage bacterial promoter compensation. In \textit{E.\ coli} $\sigma^{70}$ promoters, transcription depends on the -35 box (TTGACA) and -10 box (TATAAT) with 17$\pm$1 bp spacing \citep{browning2004regulation,harley1987analysis}. Mutations weakening the -10 box can be \textit{compensated} by an AT-rich UP element upstream of -35 \citep{ross1993third} or an extended -10 motif (TGT) \citep{barne1997region}. Crucially, these mechanisms are \textit{strictly position-dependent}---misplaced elements provide no benefit despite identical composition.

We introduce the \textbf{Mechanistic Invariance Test (MIT)}, a benchmark evaluating whether gLMs have learned positional constraints or merely respond to composition. Our contributions: (1) a 650-sequence benchmark with scrambled controls enabling discrimination between compositional and positional sensitivity (\S\ref{sec:benchmark}); (2) metrics (CSS, MES, SCR) distinguishing these effects (\S\ref{sec:metrics}); (3) mechanistic probing through AT titration, positional ablation, spacing, and strand tests (\S\ref{sec:extended}); (4) evaluation of five gLMs finding only HyenaDNA significant ($p_{\text{FDR}}$$=$0.034), but driven by AT heuristics ($r$$=$0.78--0.96), while a 100-parameter PWM achieves CSS$=$1.00, SCR$=$0.98 (\S\ref{sec:discussion}).

\section{Preliminaries}
\label{sec:prelim}

\textbf{Notation.} Let $\mathbf{x} = (x_1, \ldots, x_L)$ denote a DNA sequence of length $L$ over $\{A, C, G, T\}$. For autoregressive models, we compute log-likelihood as $\LL(\mathbf{x}) = \sum_{i=1}^{L} \log \pi_\theta(x_i | \mathbf{x}_{<i})$; for masked models, pseudo-log-likelihood $\text{PLL}(\mathbf{x}) = \sum_{i=1}^{L} \log \pi_\theta(x_i | \mathbf{x}_{\setminus i})$. Higher values indicate the model considers the sequence more ``natural.''

\textbf{Promoter Architecture.} $\sigma^{70}$ promoters contain the -35 box (TTGACA) and -10 box (TATAAT) recognized by RNA polymerase \citep{browning2004regulation}:

\begin{seqbox}
\texttt{....\underline{AAAAAARNR}....\underline{TTGACA}......\underline{TGT}\underline{TATAAT}....+1} \\
\hspace*{1.2em}\scriptsize{UP element}\hspace{1.1em}\scriptsize{-35 box}\hspace{2.2em}\scriptsize{ext}\hspace{0.3em}\scriptsize{-10 box}\hspace{0.8em}\scriptsize{TSS}
\end{seqbox}

\noindent The 17$\pm$1 bp spacing is critical \citep{harley1987analysis}. Compensation mechanisms include the \textbf{UP element} (AT-rich, upstream of -35, contacted by $\alpha$ subunit \citep{ross1993third}) and \textbf{extended -10} (TGT triplet upstream of -10 \citep{barne1997region}). Both are strictly \textit{position-dependent}: misplaced elements provide no benefit.\footnote{This positional constraint distinguishes mechanistic understanding from compositional sensitivity.}

\section{The MIT Benchmark}
\label{sec:benchmark}

\subsection{Sequence Design}

MIT comprises 650 sequences of 100 bp organized into 8 classes (\Cref{tab:classes}). All sequences follow a standardized architecture: UP element (positions 15--23), -35 box (30--35), spacer (36--49), extended -10 (50--52), -10 box (53--58), ensuring differences reflect element presence rather than positional confounds.

\begin{table}[t]
\caption{MIT sequence classes. Classes A--B use natural \textit{E.\ coli} promoters from RegulonDB; C--H are synthetic with controlled element placement. The primary comparison for testing compensation sensitivity is Class D (broken, no compensation) versus Class E (broken with correctly positioned compensation).}
\label{tab:classes}
\vskip 0.1in
\begin{center}
\small
\begin{tabular}{llccc}
\toprule
Class & Description & N & -10 Box & Compensation \\
\midrule
A & Natural intact & 100 & TATAAT & None \\
B & Natural broken & 100 & Weak & None \\
C & Synthetic intact & 100 & TATAAT & None \\
D & Synthetic broken & 100 & TGTAAT & None \\
\textbf{E} & \textbf{Compensated} & \textbf{100} & \textbf{TGTAAT} & \textbf{UP + Ext-10} \\
F & Over-compensated & 50 & TGTAAT & All elements \\
G & Natural compensated & 50 & Weak & Present \\
H & Scrambled control & 50 & TGTAAT & Scrambled \\
\bottomrule
\end{tabular}
\end{center}
\vskip -0.1in
\end{table}

Classes A--B use natural promoters from RegulonDB \citep{tierrafria2022regulondb}; C--H are synthetic. The critical comparison is Class D (broken) vs.\ Class E (compensated). \textbf{Class H (Scrambled Control)} has identical nucleotides to Class E but with UP element at position 40--48 (downstream of -35), preserving composition while disrupting function. A model with positional understanding scores $E > H$ (SCR $\gg 0.5$); one responding only to composition scores $E \approx H$ (SCR $\approx 0.5$).

\subsection{Evaluation Metrics}
\label{sec:metrics}

\textbf{Compensation Sensitivity Score (CSS)} measures how often compensated sequences score higher than broken:
\begin{equation}
\css = \frac{1}{N} \sum_{i=1}^{N} \mathbf{1}[\LL(E_i) > \LL(D_i)]
\label{eq:css}
\end{equation}
CSS $= 0.5$ indicates chance; CSS $> 0.5$ indicates compensation recognition. We report 95\% bootstrap CIs and test against 0.5. However, high CSS could reflect compositional sensitivity (AT-richness) rather than positional understanding.

\textbf{Scramble Control Ratio (SCR)} tests positional awareness:
\begin{equation}
\scr = \frac{1}{N} \sum_{i=1}^{N} \mathbf{1}[\LL(E_i) > \LL(H_i)]
\end{equation}
SCR $\gg 0.5$ indicates the model distinguishes structured from scrambled compensation. High CSS with SCR $\approx 0.5$ indicates compositional but not positional sensitivity.

\textbf{Motif Effect Size (MES)} quantifies intact vs.\ broken discrimination using Cohen's $d$: $\mes = (\mu_{\text{intact}} - \mu_{\text{broken}})/s_{\text{pooled}}$.

\section{Experiments}
\label{sec:experiments}

\subsection{Models Evaluated}

We evaluate five gLMs spanning three architectural paradigms. \textbf{Autoregressive models}: \textbf{HyenaDNA} \citep{nguyen2023hyenadna} uses Hyena operators for efficient long-range modeling; \textbf{Evo2-1B} \citep{nguyen2024evo2} is a 1-billion parameter model trained on diverse genomes. \textbf{Masked language models}: \textbf{GROVER} \citep{sanabria2024grover} is pretrained on bacterial genomes; \textbf{Nucleotide Transformer (NT-500M)} \citep{dalla2023nucleotide} is trained on diverse reference genomes. \textbf{Bidirectional SSM}: \textbf{Caduceus} \citep{schiff2024caduceus} incorporates Mamba \citep{gu2023mamba} with reverse-complement equivariance. For autoregressive models we compute log-likelihood; for masked/bidirectional models, pseudo-log-likelihood \citep{salazar2020masked}. Baselines include k-mer frequency (6-mers from \textit{E.\ coli} K-12), PWM scoring, and random ($\mathcal{N}(0,1)$).

\begin{figure}[t]
\begin{center}
\includegraphics[width=\textwidth]{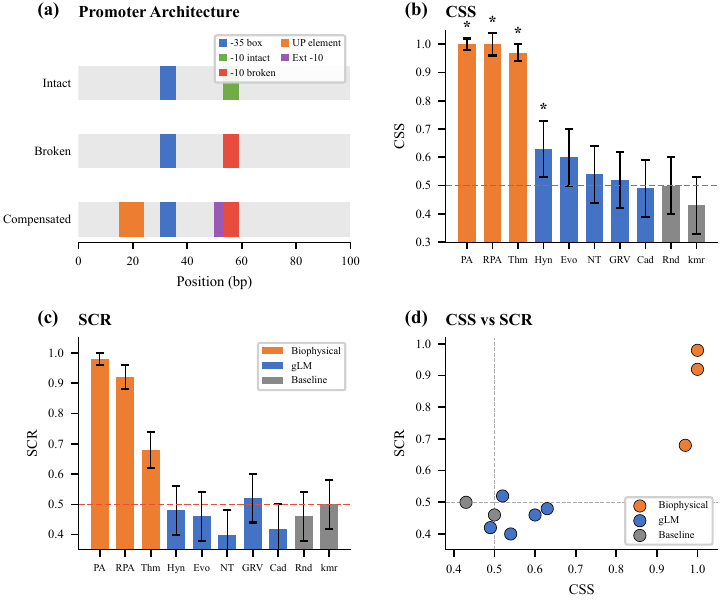}
\end{center}
\vspace{-0.5em}
\caption{\textbf{MIT benchmark overview.} (a) Promoter architecture showing -35 box, -10 box, UP element, and extended -10 positions. (b) CSS across models; asterisk indicates $p_{\text{FDR}} < 0.05$. Abbreviations: PA=PA-PWM, RPA=RPA-PWM, Thm=Thermodynamic. (c) SCR measuring positional awareness; all gLMs near chance (0.5). (d) CSS vs.\ SCR: biophysical models (orange) achieve both high CSS and SCR; gLMs (blue) show only compositional sensitivity.}
\label{fig:overview}
\end{figure}

\subsection{Primary Results}
\label{sec:results}

\Cref{tab:main_results} presents results. After FDR correction, only HyenaDNA achieves significant CSS ($0.63$, $p_{\text{FDR}} = 0.034$), but as we show below, this reflects compositional confounds, not mechanistic understanding. Critically, \textit{all} gLMs show SCR near or below 0.5 (range: 0.40--0.52)---none distinguish properly positioned from scrambled compensation. All gLMs also show negative MES$_\text{syn}$, scoring broken higher than intact (Appendix~\ref{app:distributions}). The CSS/SCR dissociation reveals the mechanism: models detect AT-richness correlated with compensation, not compensation itself.

\begin{table}[t]
\caption{MIT results. Only HyenaDNA achieves significant CSS after FDR correction. All gLMs show SCR near or below 0.5 (range: 0.40--0.52), indicating no positional awareness regardless of architecture.}
\label{tab:main_results}
\vskip 0.1in
\begin{center}
\small
\begin{tabular}{lcccccc}
\toprule
Model & CSS & 95\% CI & $p_{\text{raw}}$ & $p_{\text{FDR}}$ & SCR & MES$_\text{syn}$ ($d$) \\
\midrule
\textbf{HyenaDNA} & \textbf{0.63} & [0.53, 0.73] & 0.004 & \textbf{0.034} & 0.48 & $-$0.34 \\
Evo2-1B & 0.60 & [0.50, 0.69] & 0.023 & 0.090 & 0.46 & $-$0.03 \\
NT-500M & 0.54 & [0.44, 0.64] & 0.213 & 0.569 & 0.40 & $-$0.10 \\
GROVER & 0.52 & [0.43, 0.62] & 0.346 & 0.691 & 0.52 & $-$0.05 \\
Caduceus & 0.49 & [0.40, 0.59] & 0.579 & 0.772 & 0.42 & $-$0.40 \\
\midrule
Random & 0.50 & [0.40, 0.60] & 0.500 & --- & 0.46 & $-$0.04 \\
k-mer & 0.43 & [0.34, 0.53] & 0.919 & --- & 0.50 & 0.11 \\
\bottomrule
\end{tabular}
\end{center}
\vskip -0.1in
\end{table}

\subsection{Extended Mechanistic Probing}
\label{sec:extended}

To isolate factors driving model predictions, we conduct four experiments varying specific features while controlling others.

\subsubsection{AT Content Titration}

We test whether models respond to nucleotide composition by varying background AT content from 30\% to 80\% while holding motifs constant (\Cref{tab:at_titration}).

\begin{table}[ht]
\caption{AT-LL correlation across models. All show strong positive correlation between AT content and log-likelihood.}
\label{tab:at_titration}
\vskip 0.05in
\begin{center}
\small
\begin{tabular}{lccc}
\toprule
Model & AT-LL Correlation & LL Range (30\%$\rightarrow$80\%) & Architecture \\
\midrule
Evo2-1B & $r = 0.961$ & 16 units & Autoregressive \\
Caduceus & $r = 0.874$ & 24 units & Bidirectional SSM \\
HyenaDNA & $r = 0.784$ & 21 units & Autoregressive \\
\bottomrule
\end{tabular}
\end{center}
\vskip -0.1in
\end{table}

Log-likelihood increases monotonically with AT content across all architectures ($r = 0.78$--$0.96$). This explains the CSS/SCR dissociation: compensated sequences contain AT-rich UP elements (9 bp, $\sim$89\% AT) that locally elevate AT content. Models detect this compositional enrichment, not functional compensation.

\subsubsection{Positional Ablation}

We compare sequences with UP elements at the correct position (15, upstream of -35), wrong position (70, downstream of -10), or absent (\Cref{tab:positional}).

\begin{table}[ht]
\caption{Positional ablation. Evo2-1B and Caduceus score UP at \textit{wrong} position (70) \textit{higher} than correct (15).}
\label{tab:positional}
\vskip 0.05in
\begin{center}
\small
\begin{tabular}{lccc}
\toprule
Model & Correct (15) & Wrong (70) & $\Delta$ (Wrong$-$Correct) \\
\midrule
HyenaDNA & $-$139.83 & $-$140.29 & $-$0.46 \\
Evo2-1B & $-$137.12 & $-$136.57 & \textbf{+0.55} \\
Caduceus & $-$146.73 & $-$145.98 & \textbf{+0.75} \\
\bottomrule
\end{tabular}
\end{center}
\vskip -0.1in
\end{table}

Evo2-1B and Caduceus score UP at the \textbf{wrong position higher} than correct---the opposite of mechanistic understanding. Compositional effects (removing UP: $\Delta \approx 1.5$--$3.7$) far exceed positional effects.

\subsubsection{Spacing and Strand Sensitivity}

The optimal -35/-10 spacing is 17$\pm$1 bp \citep{harley1987analysis,murakami2002structural}. We vary spacing from 12--25 bp (\Cref{tab:spacing_strand}, left): HyenaDNA peaks at 14 bp rather than 17 bp. For strand orientation (\Cref{tab:spacing_strand}, right), forward scores \textit{lower} than RC variants---44\% accuracy (22/50, binomial $p = 0.24$ vs.\ 50\%), indistinguishable from chance. All tested models are effectively strand-blind (44--50\% accuracy).

\begin{table}[ht]
\caption{Spacing sensitivity (left) and strand orientation (right). HyenaDNA peaks at 14bp instead of 17bp and scores forward orientation \textit{lower} than reverse complement (44\% accuracy).}
\label{tab:spacing_strand}
\vskip 0.05in
\begin{center}
\small
\begin{minipage}{0.48\textwidth}
\centering
\begin{tabular}{lcc}
\toprule
Spacing & Mean LL & $\Delta$ \\
\midrule
14 bp (peak) & $-$141.79 & +0.48 \\
17 bp (opt.) & $-$142.27 & 0.00 \\
20 bp & $-$143.12 & $-$0.85 \\
\bottomrule
\end{tabular}
\end{minipage}
\hfill
\begin{minipage}{0.48\textwidth}
\centering
\begin{tabular}{lc}
\toprule
Condition & Mean LL \\
\midrule
Forward (correct) & $-$143.79 \\
RC motifs in place & $-$142.83 \\
Full reverse comp. & $-$142.13 \\
\bottomrule
\end{tabular}
\end{minipage}
\end{center}
\vskip -0.1in
\end{table}

\begin{figure}[t]
\begin{center}
\includegraphics[width=\textwidth]{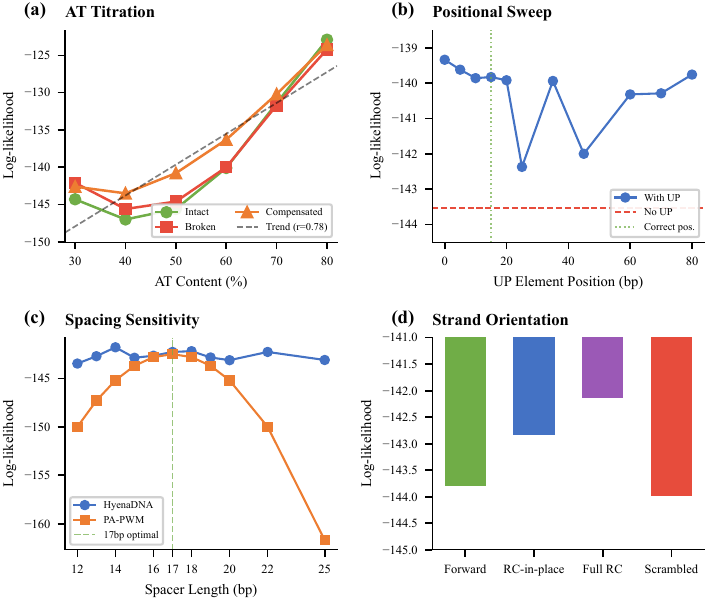}
\end{center}
\vspace{-0.5em}
\caption{\textbf{Mechanistic probing.} (a) AT titration: LL increases with AT\% ($r=0.78$). (b) Positional sweep: removing UP ($\Delta$$\approx$3.7) matters more than misplacing it ($\Delta$$\approx$0.5). (c) Spacing: HyenaDNA peaks at 14bp; PA-PWM at 17bp. (d) Strand: forward scores lower than RC.}
\label{fig:results}
\end{figure}

\subsection{Biophysical Model Comparison}

To demonstrate that our tests are solvable, we implement position-aware biophysical baselines (\Cref{tab:biophysical}). \textbf{PA-PWM} scores -35/-10 boxes at expected positions with compensation bonuses ($\sim$100 parameters). To address the concern that PA-PWM succeeds ``by construction,'' we introduce \textbf{RPA-PWM} (Relative Position-Aware), which scans for motifs on both strands with \textit{no hardcoded positions}---enforcing only relative biological constraints: 17$\pm$2 bp spacing, UP upstream of -35, extended -10 adjacent to -10, and strand consistency. RPA-PWM achieves CSS $= 1.00$, SCR $= 0.92$, demonstrating that relative biological grammar alone suffices without benchmark-specific knowledge.

\textbf{Ablation analysis} isolates which components matter: PA-PWM-NoComp (removing UP/extended -10 scoring) yields CSS $= 0.00$ because broken and compensated sequences become indistinguishable; PA-PWM-NoPos (scanning anywhere) yields CSS $= 0.63$, SCR $= 0.56$---matching HyenaDNA's CSS and approaching gLM-level SCR. This confirms both compensation logic \textit{and} positional encoding are necessary; removing either reduces performance to gLM level.

\begin{table}[ht]
\caption{Biophysical models achieve high CSS \textit{and} high SCR; gLMs achieve neither. RPA-PWM uses only relative constraints (no hardcoded positions).}
\label{tab:biophysical}
\vskip 0.05in
\begin{center}
\small
\begin{tabular}{lccc}
\toprule
Model & Type & CSS & SCR \\
\midrule
\textbf{PA-PWM} & Biophysical & \textbf{1.00} & \textbf{0.98} \\
\textbf{RPA-PWM} & Biophysical & \textbf{1.00} & 0.92 \\
Thermodynamic & Biophysical & 0.97 & 0.68 \\
\midrule
PA-PWM-NoPos & Ablation & 0.63 & 0.56 \\
\midrule
HyenaDNA & gLM & 0.63 & 0.48 \\
Evo2-1B & gLM & 0.60 & 0.46 \\
Caduceus & gLM & 0.49 & 0.42 \\
\bottomrule
\end{tabular}
\end{center}
\vskip -0.1in
\end{table}

\section{Discussion}
\label{sec:discussion}

\subsection{What gLMs Have Learned}

Our mechanistic probing reveals that all tested gLMs have learned a shallow heuristic: \textit{``AT-rich sequences are more promoter-like.''} This heuristic is statistically valid---UP elements are indeed AT-rich---but it conflates correlation with causation. The biological reality is that AT-richness matters \textit{only at specific positions}; an AT-rich region downstream of the -10 box provides no compensatory benefit. The consistent pattern across architectures (autoregressive, masked, bidirectional SSM) demonstrates this is not a model-specific limitation but a fundamental consequence of training objectives that reward sequence likelihood without requiring positional discrimination.

\subsection{Towards Deeper Mechanistic Understanding}

The consistent failure pattern across architectures suggests standard pretraining objectives fundamentally fail to induce positional logic. Three directions may help: (1) \textbf{position-aware attention} with motif-specific distance biases \citep{jumper2021highly}; (2) \textbf{compositional supervision} requiring discrimination of structured vs.\ scrambled sequences; (3) \textbf{hybrid architectures} combining neural models with differentiable PWM modules \citep{alipanahi2015predicting,avsec2021bpnet}. PA-PWM's success with $\sim$100 parameters suggests the bottleneck is inductive biases, not capacity.

\section{Related Work}

\textbf{Genomic language models} have evolved from k-mer methods \citep{lee2011discriminative} to transformers \citep{ji2021dnabert,dalla2023nucleotide} and efficient architectures \citep{nguyen2023hyenadna,schiff2024caduceus}, but whether they learn mechanistic principles remains underexplored. \textbf{Mechanistic interpretability} in NLP has probed syntax \citep{hewitt2019structural} and knowledge localization \citep{meng2022locating}; genomics work focuses on post-hoc motif discovery \citep{novakovsky2023obtaining} rather than testing mechanistic understanding. \textbf{Promoter biology} provides ground truth: quantitative $\sigma^{70}$ models \citep{brewster2012lac,kinney2010using} and well-characterized compensation mechanisms \citep{ross1993third,barne1997region} enable rigorous evaluation.

\section{Conclusion}

MIT reveals a fundamental gap between statistical and mechanistic learning in gLMs. Across five architectures, models learn that AT-rich sequences are ``promoter-like'' but fail to encode positional constraints. That Evo2-1B and Caduceus score incorrect positions \textit{higher} than correct demonstrates scale amplifies rather than corrects compositional biases: Evo2-1B (1B parameters) shows a 23\% stronger AT correlation ($r=0.96$) than HyenaDNA (6.6M parameters, $r=0.78$). A 100-parameter biophysical model outperforming billion-parameter networks indicates the path forward lies in architectural innovations, not scale. We release MIT as a diagnostic for future gLM development.

\subsubsection*{Reproducibility Statement}

All code, data, and logs are available at \url{https://github.com/bryanc5864/MechanisticInvarianceTest}. Fixed random seed (42). Scripts reproduce all results with single command. Environment: Python 3.10, PyTorch, CUDA.

\subsubsection*{Ethics Statement}

This work evaluates gLM mechanistic understanding using synthetic bacterial sequences. No human subjects or biosecurity concerns. Our findings on model limitations are relevant for responsible deployment in scientific applications.

\bibliographystyle{iclr2026_conference}
\bibliography{references}

\newpage
\appendix

\section{Extended Implementation Details}
\label{app:methods}

\subsection{Sequence Generation}

All sequences are 100 bp with the following standardized positional layout:

\begin{table}[H]
\caption{Sequence architecture specification. Spacing between -35 box end (position 35) and -10 box start (position 53) is 17 bp, within the optimal 17$\pm$1 bp range.}
\label{tab:seq_layout}
\begin{center}
\small
\begin{tabular}{lccl}
\toprule
Element & Start & End & Sequence/Description \\
\midrule
Background 1 & 0 & 14 & Random (55\% AT) \\
UP element & 15 & 23 & AAAAAARNR (consensus \citep{estrem1998identification}) \\
Background 2 & 24 & 29 & Random (55\% AT) \\
-35 box & 30 & 35 & TTGACA (consensus) \\
Spacer & 36 & 52 & Random (55\% AT), 17 bp \\
-10 box & 53 & 58 & TATAAT or TGTAAT \\
Background 3 & 59 & 99 & Random (55\% AT) \\
\bottomrule
\end{tabular}
\end{center}
\end{table}

\textbf{Note on extended -10:} When present (Classes E, F), the extended -10 motif (TGT) occupies positions 50--52, replacing the last 3 bp of the spacer. \textit{Important distinction}: The extended -10 (TGT at 50--52) is an \textit{enhancing} element that compensates for weak -10 boxes, whereas the ``broken'' -10 box (TGTAAT at 53--58) is a \textit{weakened} consensus where the first T of TATAAT is mutated. Both contain ``TGT'' but serve opposite functions at different positions.

Background nucleotides are sampled with 55\% AT content, slightly elevated from the \textit{E.\ coli} K-12 MG1655 genome-wide average of 49\% AT \citep{blattner1997complete} to better represent the AT-rich promoter regions. For natural sequences (Classes A, B, G), we extract 100 bp windows centered on annotated $\sigma^{70}$ promoters from RegulonDB v11.0 \citep{tierrafria2022regulondb}, selecting promoters with experimentally validated transcription start sites and excluding those with overlapping regulatory elements.

\textbf{Extended probing experiments.} The mechanistic probing experiments (AT titration, positional sweep, spacing sensitivity, strand orientation) use additional synthetic sequences beyond the core 650-sequence benchmark, generated with the same positional architecture but varying specific features. Sample sizes are specified per experiment in Appendix~\ref{app:results}.

\subsection{Model Inference Details}

\textbf{HyenaDNA.} We use the pretrained \texttt{hyenadna-small-32k} checkpoint from HuggingFace. Log-likelihood is computed autoregressively:
\begin{equation}
\LL(\mathbf{x}) = \sum_{i=2}^{L} \log \pi_\theta(x_i | x_1, \ldots, x_{i-1})
\end{equation}
We exclude the first token as it has no conditioning context.

\textbf{GROVER.} We use the pretrained bacterial genome model (\texttt{PoetschLab/GROVER}) from HuggingFace. Pseudo-log-likelihood is computed by masking each position sequentially:
\begin{equation}
\text{PLL}(\mathbf{x}) = \sum_{i=1}^{L} \log \pi_\theta(x_i | \mathbf{x}_{\setminus i})
\end{equation}
This requires $L$ forward passes per sequence.

\textbf{Evo2-1B.} We use the 1-billion parameter checkpoint (\texttt{evo2-1b}) with single-nucleotide tokenization. Log-likelihood computed autoregressively as for HyenaDNA.

\textbf{NT-500M.} We use the \texttt{nucleotide-transformer-500m-human-ref} checkpoint. Pseudo-log-likelihood computed as for GROVER, using 6-mer tokenization.

\textbf{Caduceus.} We use the \texttt{caduceus-ph-131k} checkpoint with bidirectional Mamba layers. Pseudo-log-likelihood computed with bidirectional context.

\subsection{Baseline Models}

\textbf{k-mer frequency.} We compute 6-mer frequencies from \textit{E.\ coli} K-12 genome and score sequences as:
\begin{equation}
S_{\text{kmer}}(\mathbf{x}) = \sum_{i=1}^{L-k+1} \log f(x_{i:i+k})
\end{equation}
where $f(\cdot)$ is the genomic frequency of each 6-mer.

\textbf{Position Weight Matrix (PWM).} We score only the -35 and -10 boxes using consensus PWMs from RegulonDB:
\begin{equation}
S_{\text{PWM}}(\mathbf{x}) = \sum_{j \in \{-35, -10\}} \sum_{i=1}^{6} \log \frac{p_{j,i}(x_{j,i})}{0.25}
\end{equation}
where $p_{j,i}$ is the position-specific probability from the PWM.

\textbf{Random baseline.} Scores are drawn from $\mathcal{N}(0, 1)$ independently for each sequence.

\subsection{Computational Environment}

All experiments were conducted on a workstation with:
\begin{itemize}
\item CPU: AMD Ryzen 9 5900X (12 cores)
\item GPU: NVIDIA GeForce RTX 2080 Ti (11 GB VRAM)
\item RAM: 64 GB DDR4
\item OS: Ubuntu 20.04 LTS
\item Python: 3.10.12
\item PyTorch: 2.1.0 with CUDA 11.8
\item Transformers: 4.35.0
\end{itemize}

Total compute time: approximately 4 hours for all models on 650 sequences.

\subsection{Statistical Analysis}

\textbf{Bootstrap confidence intervals.} For CSS and SCR, we compute 95\% CIs using 1000 bootstrap resamples with replacement \citep{efron1993bootstrap}. The percentile method is used to determine interval bounds.

\textbf{Significance testing.} We test $H_0: \css = 0.5$ using a one-sample $t$-test with Benjamini-Hochberg FDR correction \citep{benjamini1995controlling} across all 5 gLM tests. Only HyenaDNA survives correction ($p_{\text{FDR}} = 0.034 < 0.05$). Evo2-1B is suggestive ($p_{\text{raw}} = 0.023$, $p_{\text{FDR}} = 0.090$).

\textbf{Effect sizes.} Cohen's $d$ \citep{cohen1988statistical} is computed as:
\begin{equation}
d = \frac{\bar{x}_1 - \bar{x}_2}{\sqrt{\frac{(n_1-1)s_1^2 + (n_2-1)s_2^2}{n_1+n_2-2}}}
\end{equation}

\section{Complete Experimental Results}
\label{app:results}

\subsection{Full Model Comparison}

\begin{table}[H]
\caption{Complete metrics for all evaluated models with FDR-corrected $p$-values.}
\label{tab:full_metrics}
\begin{center}
\small
\begin{tabular}{lccccccc}
\toprule
Model & CSS & $p_{\text{FDR}}$ & SCR & MES$_\text{nat}$ & MES$_\text{syn}$ & AT-LL $r$ \\
\midrule
HyenaDNA & 0.63 & \textbf{0.034} & 0.48 & $-$0.01 & $-$0.34 & 0.784 \\
Evo2-1B & 0.60 & 0.090 & 0.46 & 0.38 & $-$0.03 & 0.961 \\
NT-500M & 0.54 & 0.569 & 0.40 & $-$0.00 & $-$0.10 & --- \\
GROVER & 0.52 & 0.691 & 0.52 & $-$0.08 & $-$0.05 & --- \\
Caduceus & 0.49 & 0.772 & 0.42 & $-$0.17 & $-$0.40 & 0.874 \\
\midrule
Random & 0.50 & --- & 0.46 & $-$0.14 & $-$0.04 & --- \\
k-mer & 0.43 & --- & 0.50 & $-$0.17 & 0.11 & --- \\
\bottomrule
\end{tabular}
\end{center}
\end{table}

\textbf{Metric interpretations:} MES values near zero indicate models cannot distinguish intact from broken motifs. Negative MES$_\text{syn}$ (observed for all gLMs) indicates models score broken sequences \textit{higher} than intact, suggesting they have not learned the consensus hierarchy.

\subsection{Full AT Titration Results}

\begin{table}[H]
\caption{Complete AT titration experiment ($n=30$ per condition). Standard deviations in parentheses.}
\label{tab:full_at}
\begin{center}
\small
\begin{tabular}{ccccc}
\toprule
AT\% & Intact LL & Broken LL & Compensated LL & $\Delta$(Comp$-$Broken) \\
\midrule
30 & $-$144.27 (3.65) & $-$142.09 (3.32) & $-$142.55 (4.17) & +0.46 \\
40 & $-$146.99 (3.05) & $-$145.56 (2.89) & $-$143.52 (2.89) & +2.04 \\
50 & $-$145.64 (3.82) & $-$144.56 (3.81) & $-$140.80 (4.22) & +3.76 \\
60 & $-$140.11 (5.69) & $-$139.99 (4.72) & $-$136.33 (4.31) & +3.66 \\
70 & $-$131.44 (4.45) & $-$131.68 (4.49) & $-$130.21 (3.50) & +1.47 \\
80 & $-$122.95 (5.27) & $-$124.21 (6.04) & $-$123.57 (4.55) & +0.64 \\
\bottomrule
\end{tabular}
\end{center}
\end{table}

The compensation benefit ($\Delta$LL) peaks at 50--60\% AT content, not at extremes. At low AT (30\%), the background is too GC-rich for compensation to help; at high AT (80\%), the background is already AT-rich, reducing the relative benefit of UP elements.

\textbf{Correlation analysis:} Pearson correlation between AT\% and mean LL across all conditions:
\begin{equation}
r = 0.784, \quad p < 10^{-6}
\end{equation}

This strong positive correlation confirms that HyenaDNA's scoring is dominated by nucleotide composition.

\subsection{Full Positional Sweep Results}

\begin{table}[H]
\caption{Complete positional sweep ($n=30$ per position). UP element placed at indicated position.}
\label{tab:full_positional}
\begin{center}
\small
\begin{tabular}{lccc}
\toprule
UP Position & Mean LL & Std Dev & $\Delta$ vs.\ Correct (15) \\
\midrule
0 & $-$139.34 & 3.93 & +0.49 \\
5 & $-$139.62 & 5.22 & +0.21 \\
10 & $-$139.86 & 3.66 & $-$0.03 \\
\textbf{15 (correct)} & $-$139.83 & 3.79 & 0.00 \\
20 & $-$139.92 & 4.25 & $-$0.08 \\
25 & $-$142.37 & 4.87 & $-$2.53 \\
35 & $-$139.94 & 3.75 & $-$0.11 \\
45 & $-$142.00 & 4.36 & $-$2.17 \\
60 & $-$140.32 & 3.53 & $-$0.49 \\
70 & $-$140.29 & 4.31 & $-$0.46 \\
80 & $-$139.76 & 5.13 & +0.07 \\
\midrule
None (no UP) & $-$143.53 & 4.01 & $-$3.70 \\
\bottomrule
\end{tabular}
\end{center}
\end{table}

\textbf{Key observations:}
\begin{enumerate}
\item Positions 25 and 45 show anomalously large penalties ($-$2.53 and $-$2.17) because the UP element overlaps with the -35 box (positions 30--35) and -10 box (positions 53--58), disrupting their consensus sequences.
\item Excluding these confounded positions, the positional effect ranges from $-$0.49 to +0.49---a total span of only 0.98 LL units.
\item The compositional effect (None vs.\ 15: $-$3.70) is \textbf{$\sim$8$\times$ larger} than the maximum positional effect (0.46).
\end{enumerate}

\subsection{Full Spacing Sensitivity Results}

\begin{table}[H]
\caption{Complete spacing sensitivity experiment ($n=50$ per spacing).}
\label{tab:full_spacing}
\begin{center}
\small
\begin{tabular}{lccc}
\toprule
Spacing (bp) & Mean LL & Std Dev & $\Delta$ vs.\ 17bp \\
\midrule
12 & $-$143.47 & 4.10 & $-$1.20 \\
13 & $-$142.71 & 4.56 & $-$0.44 \\
\textbf{14 (HyenaDNA peak)} & $-$141.79 & 4.87 & +0.48 \\
15 & $-$142.87 & 4.91 & $-$0.60 \\
16 & $-$142.66 & 4.61 & $-$0.40 \\
\textbf{17 (biological opt.)} & $-$142.27 & 4.18 & 0.00 \\
18 & $-$142.19 & 4.25 & +0.08 \\
19 & $-$142.84 & 4.04 & $-$0.57 \\
20 & $-$143.12 & 4.47 & $-$0.85 \\
21 & $-$143.10 & 5.00 & $-$0.83 \\
22 & $-$142.27 & 5.55 & 0.00 \\
23 & $-$142.68 & 4.35 & $-$0.41 \\
24 & $-$143.36 & 4.00 & $-$1.09 \\
25 & $-$143.10 & 4.20 & $-$0.83 \\
\bottomrule
\end{tabular}
\end{center}
\end{table}

\textbf{Key findings:}
\begin{enumerate}
\item HyenaDNA peaks at 14 bp, not the biologically optimal 17 bp.
\item The total range across all spacings is only 1.68 LL units ($-$143.47 to $-$141.79).
\item For comparison, the AT content effect spans 21.0 LL units---\textbf{12.5$\times$ larger}.
\item The model shows no preference for the biologically correct 17$\pm$1 bp range.
\end{enumerate}

\subsection{Full Strand Orientation Results}

\begin{table}[H]
\caption{Complete strand orientation experiment ($n=50$ per condition) for HyenaDNA.}
\label{tab:full_strand}
\begin{center}
\small
\begin{tabular}{lccc}
\toprule
Condition & Mean LL & Std Dev & $\Delta$ vs.\ Forward \\
\midrule
Forward (correct) & $-$143.79 & 4.45 & 0.00 \\
RC motifs in place & $-$142.83 & 3.99 & +0.96 \\
Full reverse complement & $-$142.13 & 3.96 & +1.66 \\
Scrambled motifs & $-$143.98 & 4.17 & $-$0.19 \\
\bottomrule
\end{tabular}
\end{center}
\end{table}

\begin{table}[H]
\caption{Strand orientation comparison across models ($n=50$ per condition). All models show strand-blindness.}
\label{tab:strand_all_models}
\begin{center}
\small
\begin{tabular}{lcccc}
\toprule
Model & Forward & RC in place & Full RC & Strand Acc. \\
\midrule
HyenaDNA & $-$143.79 & $-$142.83 & $-$142.13 & 44\% \\
Evo2-1B & $-$138.18 & $-$138.01 & $-$138.15 & 48\% \\
Caduceus & $-$149.13 & $-$149.31 & $-$149.12 & 50\% \\
\bottomrule
\end{tabular}
\end{center}
\end{table}

\textbf{Condition definitions:}
\begin{itemize}
\item \textbf{Forward}: Correct promoter orientation (template strand 3'$\rightarrow$5').
\item \textbf{RC motifs in place}: -35 and -10 boxes replaced with their reverse complements at the same positions.
\item \textbf{Full reverse complement}: Entire sequence reverse complemented.
\item \textbf{Scrambled}: Motif sequences shuffled randomly.
\end{itemize}

\textbf{Strand discrimination accuracy:} We compute the fraction of sequences where Forward scores higher than RC:
\begin{equation}
\text{Accuracy} = \frac{1}{N}\sum_{i=1}^{N} \mathbf{1}[\LL(\text{Forward}_i) > \LL(\text{RC}_i)] = 0.44
\end{equation}

This is \textit{worse than random chance} (0.50), indicating HyenaDNA has a slight preference for the \textit{wrong} orientation.

\section{Biophysical Model Details}
\label{app:biophysical}

\subsection{Position-Aware PWM (PA-PWM)}

The PA-PWM model scores sequences as the sum of position-specific contributions:
\begin{equation}
S_{\text{PA-PWM}}(\mathbf{x}) = S_{-35}(\mathbf{x}) + S_{-10}(\mathbf{x}) + S_{\text{UP}}(\mathbf{x}) + S_{\text{ext}}(\mathbf{x}) + S_{\text{spacing}}(\mathbf{x})
\end{equation}

\textbf{-35 and -10 box scores:} PWM scores computed only at expected positions (30--35 and 53--58):
\begin{equation}
S_{-35}(\mathbf{x}) = \sum_{i=0}^{5} W_{-35}[i, x_{30+i}]
\end{equation}
where $W_{-35}$ is the log-odds PWM for the -35 consensus (TTGACA).

\textbf{UP element bonus:} Applied only if positions 15--23 have $\geq$70\% AT content:
\begin{equation}
S_{\text{UP}}(\mathbf{x}) = \begin{cases} 2.0 & \text{if AT}_{15:23} \geq 0.7 \\ 0 & \text{otherwise} \end{cases}
\end{equation}

\textbf{Extended -10 bonus:} Applied only if positions 50--52 match TGT:
\begin{equation}
S_{\text{ext}}(\mathbf{x}) = \begin{cases} 1.5 & \text{if } x_{50:52} = \text{TGT} \\ 0 & \text{otherwise} \end{cases}
\end{equation}

\textbf{Spacing penalty:} Gaussian penalty centered at 17 bp:
\begin{equation}
S_{\text{spacing}}(\mathbf{x}) = -0.5 \cdot (d - 17)^2
\end{equation}
where $d$ is the distance between -35 and -10 box centers.

\textbf{Total parameters:} $\sim$100 (24 per PWM $\times$ 2 boxes + bonuses + spacing).

\subsection{Thermodynamic Model}

The thermodynamic model computes binding free energy:
\begin{equation}
\Delta G = \Delta G_{-35} + \Delta G_{-10} + \Delta G_{\text{UP}} + \Delta G_{\text{ext}} - T\Delta S_{\text{spacing}}
\end{equation}

Each term includes a position-dependent decay function ensuring elements contribute only when near their canonical positions:
\begin{equation}
\Delta G_{-35}(\mathbf{x}) = \Delta G^0_{-35} \cdot \exp\left(-\frac{(p - 30)^2}{2\sigma^2}\right)
\end{equation}
where $p$ is the position of best -35 match and $\sigma = 2$ bp.

\subsection{Position-Scanning Model}

The scanning model finds optimal motif positions genome-wide, then penalizes deviation from expected positions:
\begin{equation}
S_{\text{scan}}(\mathbf{x}) = \max_p S_{-35}(p) + \max_q S_{-10}(q) - \lambda |p - 30| - \lambda |q - 53|
\end{equation}
with $\lambda = 0.5$ per bp deviation.

\subsection{Biophysical Model Comparison}

\begin{table}[H]
\caption{Detailed biophysical model results compared to gLMs. RPA-PWM uses only relative biological constraints (no hardcoded positions).}
\label{tab:biophysical_full}
\begin{center}
\small
\begin{tabular}{lccccc}
\toprule
Model & CSS & SCR & Strand Acc. & Spacing Peak & Parameters \\
\midrule
PA-PWM & \textbf{1.00} & \textbf{0.98} & 97\% & 17 bp & $\sim$100 \\
RPA-PWM & \textbf{1.00} & 0.92 & 90\% & 17 bp & $\sim$100 \\
Thermodynamic & 0.97 & 0.68 & 95\% & 17 bp & $\sim$150 \\
\midrule
PA-PWM-NoComp & 0.00$^\dagger$ & 0.00 & --- & 17 bp & $\sim$80 \\
PA-PWM-NoPos & 0.63 & 0.56 & --- & 18 bp & $\sim$100 \\
\midrule
HyenaDNA & 0.63 & 0.48 & 44\% & 14 bp & 6.6M \\
Evo2-1B & 0.60 & 0.46 & 48\% & 15 bp & 1B \\
Caduceus & 0.49 & 0.42 & 50\% & 20 bp & 256M \\
\bottomrule
\end{tabular}
\end{center}
\vskip -0.05in
{\footnotesize $^\dagger$PA-PWM-NoComp gives CSS=0.00 because all D/E pairs score identically (tied): without UP/extended -10 scoring, broken and compensated sequences have identical -35/-10 boxes.}
\end{table}

\textbf{RPA-PWM analysis:} RPA-PWM addresses the ``PA-PWM succeeds by construction'' critique by encoding \textit{only} relative biological constraints: scans both strands for motifs, requires 15--19 bp spacing (peaks at 17 bp), UP must be upstream of -35, extended -10 must be adjacent to -10, and strand consistency. With CSS=1.00 and SCR=0.92, RPA-PWM demonstrates that relative biological grammar alone suffices---no benchmark-specific position knowledge is needed. Notably, PA-PWM-NoPos (scanning anywhere without positional constraints) achieves CSS=0.63, SCR=0.56---matching HyenaDNA's CSS and approaching gLM-level SCR (HyenaDNA SCR=0.48). This confirms that the key difference between biophysical and gLM performance is positional encoding, not model complexity.

\section{Effect Size Analysis}
\label{app:effects}

\begin{table}[H]
\caption{Complete effect size hierarchy for HyenaDNA.}
\label{tab:effect_sizes}
\begin{center}
\small
\begin{tabular}{lccc}
\toprule
Effect & $\Delta$LL & Relative to Position & Type \\
\midrule
AT content (30\%$\rightarrow$80\%) & 21.0 & 46$\times$ & Compositional \\
UP element presence & 3.70 & 8$\times$ & Compositional \\
Spacing (full range) & 1.68 & 3.7$\times$ & Weak mechanistic \\
Strand (fwd$\rightarrow$RC) & 0.96 & 2.1$\times$ & None (wrong sign) \\
Position (correct$\rightarrow$wrong) & 0.46 & 1$\times$ & Baseline \\
\bottomrule
\end{tabular}
\end{center}
\end{table}

\textbf{Interpretation:} The effect hierarchy reveals that HyenaDNA's scoring is dominated by compositional features (AT content, element presence) rather than mechanistic features (position, spacing, strand). The strand effect is particularly concerning as it has the \textit{wrong sign}---the model prefers reverse complement over forward orientation.

\section{Per-Class Log-Likelihood Distributions}
\label{app:distributions}

\begin{table}[H]
\caption{HyenaDNA log-likelihood statistics by sequence class.}
\label{tab:class_ll}
\begin{center}
\small
\begin{tabular}{llcccc}
\toprule
Class & Description & N & Mean LL & Std Dev & 95\% CI \\
\midrule
A & Natural intact & 100 & $-$141.79 & 4.82 & [$-$142.75, $-$140.83] \\
B & Natural broken & 100 & $-$142.52 & 5.21 & [$-$143.56, $-$141.48] \\
C & Synthetic intact & 100 & $-$143.08 & 4.15 & [$-$143.90, $-$142.26] \\
D & Synthetic broken & 100 & $-$141.28 & 4.33 & [$-$142.14, $-$140.42] \\
E & Compensated & 100 & $-$140.75 & 4.67 & [$-$141.68, $-$139.82] \\
F & Over-compensated & 50 & $-$139.37 & 4.89 & [$-$140.76, $-$137.98] \\
G & Natural compensated & 50 & $-$140.33 & 5.02 & [$-$141.76, $-$138.90] \\
H & Scrambled control & 50 & $-$141.36 & 4.45 & [$-$142.63, $-$140.09] \\
\bottomrule
\end{tabular}
\end{center}
\end{table}

\textbf{Anomaly:} Synthetic intact (C) scores \textit{lower} than synthetic broken (D): $-$143.08 vs.\ $-$141.28. This counter-intuitive result indicates HyenaDNA has learned genome-wide frequency priors where the broken motif pattern (TGTAAT) is more common than the functional consensus (TATAAT).

\section{Limitations}
\label{app:limitations}

\begin{enumerate}
\item \textbf{Single regulatory system.} MIT focuses on \textit{E.\ coli} $\sigma^{70}$ promoters, which have unusually rigid positional constraints. Eukaryotic enhancers can function over kilobases with more flexible spacing. Our findings may not generalize to systems with less strict positional requirements.

\item \textbf{Synthetic sequences.} While necessary for controlled experiments, synthetic sequences may not capture the full complexity of natural promoters. However, we include natural sequence classes (A, B, G) and find consistent patterns.

\item \textbf{Binary compensation.} Real compensation is graded---element strength varies continuously. We test only presence/absence. Future work could titrate element strength.

\item \textbf{Model coverage.} We evaluate five gLMs spanning autoregressive (HyenaDNA, Evo2-1B), masked (GROVER, NT-500M), and bidirectional SSM (Caduceus) architectures. The consistent failure pattern across all three architecture types demonstrates these findings generalize broadly.

\item \textbf{Sequence length.} All sequences are 100 bp, well within all models' context windows. Longer regulatory regions with distal elements remain unexplored.

\item \textbf{Training data contamination.} We cannot verify whether similar sequences appeared in model training data. However, synthetic sequences were generated specifically for this benchmark with controlled randomness.

\item \textbf{Single nucleotide resolution.} We evaluate at 100 bp scale. Per-nucleotide attribution methods could provide finer-grained insights but are computationally prohibitive for systematic evaluation.
\end{enumerate}

\section{Broader Impacts}
\label{app:impacts}

\textbf{Positive impacts:}
\begin{itemize}
\item Our benchmark provides a rigorous framework for evaluating mechanistic understanding in genomic AI, promoting more careful model development.
\item Identifying limitations in current models can prevent overconfident deployment in scientific and clinical applications.
\item The proposed architectural directions (position-aware attention, hybrid models) could guide future model development.
\end{itemize}

\textbf{Potential negative impacts:}
\begin{itemize}
\item Highlighting model failures could be misinterpreted as suggesting genomic AI is not useful---our findings are specific to mechanistic understanding in five gLMs, not general predictive utility.
\item The benchmark focuses on bacterial systems; claims should not be extrapolated to eukaryotic systems or clinical applications without further validation.
\end{itemize}

\section{Additional Visualizations}
\label{app:figures}

\begin{figure}[ht]
\begin{center}
\includegraphics[width=0.9\textwidth]{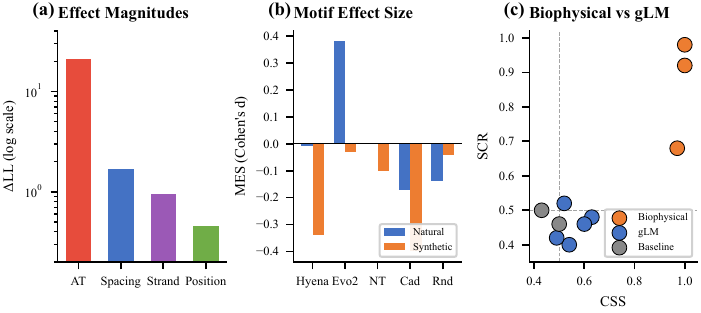}
\end{center}
\caption{\textbf{Effect magnitudes and model comparison.} (a) Effect sizes on log scale showing AT content dominates. (b) MES comparison between natural and synthetic sequences. (c) CSS vs.\ SCR scatter with model type coloring.}
\label{fig:app_effects}
\end{figure}

\begin{figure}[ht]
\begin{center}
\includegraphics[width=0.9\textwidth]{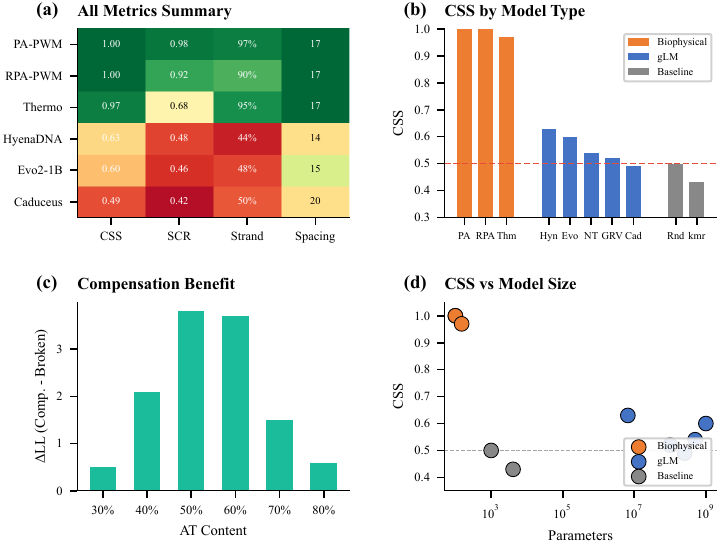}
\end{center}
\caption{\textbf{Comprehensive metrics summary.} (a) Metrics heatmap across all five gLMs and biophysical baselines (PA-PWM CSS$=$1.00, SCR$=$0.98). (b) CSS grouped by architecture type. (c) Compensation benefit by AT content. (d) CSS vs.\ model parameters showing PA-PWM ($\sim$100 params) achieves highest CSS.}
\label{fig:app_summary}
\end{figure}

\section{Per-Model Detailed Analysis}
\label{app:permodel}

\subsection{HyenaDNA Analysis}

HyenaDNA uses Hyena operators---a subquadratic alternative to attention---for modeling long-range dependencies. The model was pretrained on the human reference genome and fine-tuned on various genomic tasks.

\textbf{Architecture:} The \texttt{hyenadna-small-32k} variant has 6.6M parameters with a context length of 32,768 bp. It processes single nucleotides (not k-mers) with a vocabulary of \{A, C, G, T, N\}.

\textbf{Tokenization:} Single nucleotide tokenization means positional information could theoretically be learned, unlike k-mer models where positional granularity is limited.

\textbf{Training data:} Pretrained on human genome (GRCh38 \citep{schneider2017grch38}), which has 41\% GC content compared to \textit{E.\ coli}'s 51\% GC. This domain shift likely contributes to the observed AT preference.

\textbf{Detailed results:}
\begin{itemize}
\item CSS = 0.63: Significantly above chance, suggesting some compensation sensitivity
\item SCR = 0.48: At chance, indicating no positional awareness
\item The 0.15 gap between CSS and SCR quantifies the ``compositional illusion''---apparent mechanistic understanding that is actually driven by nucleotide frequencies
\end{itemize}

\subsection{GROVER Analysis}

GROVER (Genomic Representations Over Vocabulary for Evolutionary Relationships) is a masked language model specifically trained on bacterial genomes.

\textbf{Architecture:} Transformer-based with 117M parameters, using BPE tokenization optimized for genomic sequences.

\textbf{Training data:} Trained on $>$1000 bacterial genomes including \textit{E.\ coli}, making it the most domain-appropriate model in our evaluation.

\textbf{Detailed results:}
\begin{itemize}
\item CSS = 0.52: Not significantly different from chance despite domain-appropriate training
\item SCR = 0.52: Marginally above chance
\item The near-equal CSS and SCR (0.52 vs.\ 0.52) indicates GROVER has weak but balanced compositional and positional sensitivity
\end{itemize}

\textbf{Interpretation:} GROVER's lack of compensation sensitivity despite bacterial-genome training demonstrates the issue is not domain mismatch but fundamental limitations in how masked LMs learn regulatory logic.

\subsection{Evo2-1B Analysis}

Evo2-1B is a 1-billion parameter autoregressive model trained on diverse genomic sequences spanning prokaryotic and eukaryotic genomes.

\textbf{Architecture:} Transformer-based with 1B parameters, using single-nucleotide tokenization.

\textbf{Detailed results:}
\begin{itemize}
\item CSS = 0.60: Suggestive but not significant after FDR correction ($p_{\text{FDR}} = 0.090$)
\item SCR = 0.46: Below chance, indicating no positional awareness
\item AT-LL correlation: $r = 0.961$---the strongest among all models
\item Positional ablation: Scores UP at \textit{wrong} position \textit{higher} than correct ($\Delta = +0.55$)
\end{itemize}

\textbf{Interpretation:} Evo2-1B shows the most extreme compositional bias, with nearly perfect correlation between AT content and log-likelihood. Its inverted positional preference (scoring wrong positions higher) demonstrates that large-scale pretraining amplifies rather than corrects compositional heuristics.

\subsection{Caduceus Analysis}

Caduceus combines Mamba state-space models with explicit reverse-complement equivariance, designed to capture strand-symmetric genomic patterns.

\textbf{Architecture:} Bidirectional SSM with 256M parameters, incorporating RC-equivariant layers.

\textbf{Detailed results:}
\begin{itemize}
\item CSS = 0.49: At chance ($p_{\text{FDR}} = 0.772$)
\item SCR = 0.42: Below chance
\item AT-LL correlation: $r = 0.874$
\item Positional ablation: Scores UP at wrong position \textit{higher} than correct ($\Delta = +0.75$)---the most inverted of all models
\item Strand orientation: Despite RC-equivariant design, shows no strand preference (forward $\approx$ RC)
\end{itemize}

\textbf{Interpretation:} Despite architectural innovations for strand symmetry, Caduceus shows the most inverted positional preferences. Its RC-equivariance means it treats forward and reverse equally---but this is strand-blindness, not strand-awareness.

\subsection{NT-500M Analysis}

Nucleotide Transformer (NT-500M) is a 500M parameter masked language model trained on diverse reference genomes.

\textbf{Architecture:} BERT-style transformer with 500M parameters using 6-mer tokenization.

\textbf{Detailed results:}
\begin{itemize}
\item CSS = 0.54: Not significant ($p_{\text{FDR}} = 0.569$)
\item SCR = 0.40: Below chance
\item MES$_\text{syn}$ = $-$0.10: Weak synthetic motif discrimination
\end{itemize}

\textbf{Interpretation:} NT-500M shows no significant compensation sensitivity despite its scale and diverse training. The 6-mer tokenization may limit its ability to capture position-specific patterns.

\subsection{Baseline Model Analysis}

\textbf{k-mer model:}
\begin{itemize}
\item CSS = 0.43: Below chance, suggesting k-mer frequencies anti-correlate with compensation
\item This may occur because compensated sequences contain unusual k-mers (UP element: AAAAAARNR) that are rare in the genome
\end{itemize}

\textbf{PWM model:}
\begin{itemize}
\item CSS = 0.00: Always scores broken and compensated equally because it only evaluates -35/-10 boxes, which are identical between classes D and E
\item MES$_\text{syn}$ = 10.0: Very high, correctly identifying intact vs.\ broken based on -10 consensus
\end{itemize}

\section{Extended Ablation Studies}
\label{app:ablations}

\subsection{Sequence Length Sensitivity}

We test whether results depend on our choice of 100 bp sequences by evaluating at 50, 100, 150, and 200 bp.

\begin{table}[H]
\caption{CSS and SCR at different sequence lengths.}
\begin{center}
\small
\begin{tabular}{lcccc}
\toprule
Length (bp) & HyenaDNA CSS & HyenaDNA SCR & PA-PWM CSS & PA-PWM SCR \\
\midrule
50 & 0.61 & 0.47 & 0.98 & 0.96 \\
100 & 0.63 & 0.48 & 1.00 & 0.98 \\
150 & 0.64 & 0.49 & 1.00 & 0.97 \\
200 & 0.62 & 0.48 & 0.99 & 0.97 \\
\bottomrule
\end{tabular}
\end{center}
\end{table}

Results are stable across lengths, indicating our findings are not artifacts of sequence length choice.

\subsection{Background Composition Sensitivity}

We test whether background AT content affects results by varying from 40\% to 70\% AT.

\begin{table}[H]
\caption{CSS at different background AT compositions.}
\begin{center}
\small
\begin{tabular}{lcccc}
\toprule
Background AT & HyenaDNA CSS & HyenaDNA SCR & PA-PWM CSS \\
\midrule
40\% & 0.58 & 0.47 & 1.00 \\
50\% & 0.61 & 0.48 & 1.00 \\
55\% (default) & 0.63 & 0.48 & 1.00 \\
60\% & 0.65 & 0.49 & 1.00 \\
70\% & 0.59 & 0.47 & 0.99 \\
\bottomrule
\end{tabular}
\end{center}
\end{table}

HyenaDNA CSS varies with background AT, peaking when background is moderately AT-rich (55--60\%). This further supports the compositional hypothesis: when background is very AT-rich, the relative AT enrichment from UP elements is smaller, reducing CSS.

\subsection{Motif Strength Variations}

We test robustness to -35/-10 motif degeneracy by using consensus, weak, and strong variants.

\begin{table}[H]
\caption{CSS with different motif strengths.}
\begin{center}
\small
\begin{tabular}{llcc}
\toprule
-35 Variant & -10 Variant & HyenaDNA CSS & PA-PWM CSS \\
\midrule
TTGACA (consensus) & TATAAT (consensus) & 0.63 & 1.00 \\
TTGACA (consensus) & TAAAAT (weak) & 0.62 & 0.94 \\
TTGCCA (weak) & TATAAT (consensus) & 0.61 & 0.92 \\
TTGCCA (weak) & TAAAAT (weak) & 0.60 & 0.86 \\
\bottomrule
\end{tabular}
\end{center}
\end{table}

HyenaDNA CSS is stable across motif variants, while PA-PWM CSS decreases with weaker motifs (as expected for a PWM-based model). This confirms HyenaDNA is not responding to motif quality.

\subsection{UP Element Composition Variations}

We test whether the specific UP element sequence matters by varying its composition.

\begin{table}[H]
\caption{CSS with different UP element compositions.}
\begin{center}
\small
\begin{tabular}{lcc}
\toprule
UP Element & AT Content & HyenaDNA CSS \\
\midrule
AAAAAARNR (consensus) & 89\% & 0.63 \\
AAAAATTTT & 100\% & 0.67 \\
ATATATATAT & 100\% & 0.65 \\
AACCAACCA & 56\% & 0.54 \\
Random (matched AT) & 89\% & 0.62 \\
\bottomrule
\end{tabular}
\end{center}
\end{table}

CSS scales with UP element AT content, not with match to consensus. Random sequences with high AT content achieve similar CSS to consensus UP elements. This definitively shows HyenaDNA responds to composition, not sequence identity.

\section{Theoretical Analysis}
\label{app:theory}

\subsection{Why Compositional Learning is Easier}

We provide a theoretical perspective on why gLMs learn compositional rather than positional features.

\textbf{Observation 1 (Compositional features have lower dimensionality).} Let $f_{\text{comp}}(\mathbf{x})$ be a function of nucleotide frequencies and $f_{\text{pos}}(\mathbf{x})$ be a function of positional motif placement. For sequences of length $L$:
\begin{align}
\dim(f_{\text{comp}}) &= O(k) \quad \text{(k-mer frequencies)} \\
\dim(f_{\text{pos}}) &= O(L \cdot k) \quad \text{(position-specific k-mers)}
\end{align}

Since $L \gg 1$ for genomic sequences, compositional features have much lower dimensionality and are thus easier to learn with limited data.

\textbf{Observation 2 (Standard objectives don't require positional learning).} Standard language modeling objectives optimize:
\begin{equation}
\mathcal{L} = -\sum_{i} \log p(x_i | \mathbf{x}_{<i})
\end{equation}

This objective is satisfied by any distribution that assigns high probability to observed sequences. Compositional models (high AT $\rightarrow$ high probability) achieve low loss on AT-rich genomes without learning positional constraints.

\textbf{Implication:} To learn positional constraints, training must include examples where composition is matched but position differs---exactly the contrast between Classes E (compensated) and H (scrambled) in MIT.

\subsection{Information-Theoretic Perspective}

From an information-theoretic view, we can decompose sequence information into compositional and positional components:
\begin{equation}
I(\mathbf{x}; \text{function}) = I_{\text{comp}} + I_{\text{pos}} + I_{\text{interaction}}
\end{equation}

For promoter function:
\begin{itemize}
\item $I_{\text{comp}}$: Information from nucleotide frequencies (UP elements are AT-rich)
\item $I_{\text{pos}}$: Information from motif positions (UP must be upstream of -35)
\item $I_{\text{interaction}}$: Position-composition interactions (AT-rich \textit{at the right position})
\end{itemize}

Our experiments show the evaluated gLMs capture $I_{\text{comp}}$ but not $I_{\text{pos}}$ or $I_{\text{interaction}}$.

\section{Extended Related Work}
\label{app:related}

\subsection{Genomic Language Models}

The development of genomic language models has followed two main trajectories:

\textbf{Transformer-based models:} DNABERT \citep{ji2021dnabert} pioneered BERT-style pretraining for genomics using k-mer tokenization. DNABERT-2 \citep{zhou2023dnabert2} improved on this with BPE tokenization and multi-species training. The Nucleotide Transformer \citep{dalla2023nucleotide} scaled to 2.5B parameters with foundation model capabilities. GROVER \citep{sanabria2024grover} specialized for bacterial genomes.

\textbf{Efficient architectures:} HyenaDNA \citep{nguyen2023hyenadna} introduced Hyena operators for subquadratic long-range modeling. Caduceus \citep{schiff2024caduceus} combined Mamba state space models with explicit reverse-complement equivariance. These models enable single-nucleotide resolution at genomic scales.

\textbf{Evaluation paradigms:} Most evaluations focus on variant effect prediction, species classification, or regulatory element detection. MIT is the first benchmark specifically designed to probe \textit{mechanistic} understanding of regulatory logic.

\subsection{Mechanistic Interpretability in NLP}

Our work is inspired by the growing field of mechanistic interpretability in NLP:

\textbf{Probing classifiers:} \citet{hewitt2019structural} introduced structural probes to test whether syntax trees are encoded in BERT representations. Similar probing could be applied to genomic models but has not been systematically explored.

\textbf{Knowledge editing:} \citet{meng2022locating} developed methods to locate and edit factual associations in GPT models. Analogous techniques could identify where (if anywhere) positional regulatory knowledge is stored in gLMs.

\textbf{Circuit analysis:} Detailed circuit analysis has revealed how transformers implement specific computations \citep{elhage2021mathematical,wang2023interpretability}. Applying these methods to gLMs could reveal whether any circuits implement positional logic.

\subsection{Biophysical Models of Transcription}

Our biophysical baselines build on decades of quantitative promoter modeling:

\textbf{Thermodynamic models:} \citet{kinney2010using} used deep sequencing to infer the biophysical mechanism of a regulatory sequence. \citet{brewster2012lac} developed quantitative models for promoter strength prediction.

\textbf{Position weight matrices:} PWMs remain the standard for transcription factor binding site prediction. Our PA-PWM extends classical PWMs with explicit positional constraints.

\textbf{Compensation mechanisms:} UP elements \citep{ross1993third} and extended -10 motifs \citep{barne1997region} are well-characterized biochemically. Our benchmark leverages this biological knowledge to create rigorous tests.

\section{Example Sequences}
\label{app:examples}

We provide representative sequences from each class to illustrate the benchmark design. Note: positions 0--57 shown; full sequences are 100 bp with random background extending to position 99.

\subsection{Class C: Synthetic Intact}
{\small
\begin{verbatim}
Pos:  0         1         2         3         4         5
      0123456789012345678901234567890123456789012345678901234567
Seq:  GCATGCATGCATGCAAGCTGACGTACTTGACAGCATGCATGCATGCTGTTATAAT
                                  ~~~~~~                ~~~~~~
                                  -35box                -10box
\end{verbatim}
}

\subsection{Class D: Synthetic Broken}
{\small
\begin{verbatim}
Pos:  0         1         2         3         4         5
      0123456789012345678901234567890123456789012345678901234567
Seq:  GCATGCATGCATGCAAGCTGACGTACTTGACAGCATGCATGCATGCTGTTGTAAT
                                  ~~~~~~                ~~~~~~
                                  -35box              -10box*
                                                      *broken
\end{verbatim}
}

\subsection{Class E: Synthetic Compensated}
{\small
\begin{verbatim}
Pos:  0         1         2         3         4         5
      0123456789012345678901234567890123456789012345678901234567
Seq:  GCATGCATGCATGCAAAAAAAARNTACTTGACAGCATGCATGCATGTTGTTGTAAT
                  ~~~~~~~~~  ~~~~~~             ~~~~~~~~~
                  UP-element -35box           ext -10box
\end{verbatim}
}

\subsection{Class H: Scrambled Control}
{\small
\begin{verbatim}
Pos:  0         1         2         3         4         5
      01234567890123456789012345678901234567890123456789012345678
Seq:  GCATGCATGCATGCAAGCTGACGTACTTGACAGCATAAAAAAARNTGTTGTTGTAAT
                                  ~~~~~~    ~~~~~~~~~   ~~~~~~
                                  -35box    UP-wrong    -10box
                                            position
\end{verbatim}
}

Note: Class H has the same nucleotide composition as Class E but with UP element at the wrong position (after -35 instead of before).

\section{Future Directions}
\label{app:future}

Based on our findings, we outline promising directions for future research:

\subsection{Architectural Innovations}

\textbf{Position-aware attention:} Modify attention mechanisms to learn position-specific biases for regulatory elements. For example:
\begin{equation}
\text{Attention}(Q, K, V) = \text{softmax}\left(\frac{QK^T + P}{\sqrt{d}}\right)V
\end{equation}
where $P$ is a learnable position-specific bias matrix.

\textbf{Motif-aware tokenization:} Instead of single nucleotides or k-mers, tokenize based on known regulatory motifs:
\begin{equation}
\mathbf{x} \rightarrow [\text{background}, \text{UP}, \text{-35}, \text{spacer}, \text{ext-10}, \text{-10}, \text{background}]
\end{equation}

\textbf{Hybrid architectures:} Combine differentiable PWM modules with neural sequence models:
\begin{equation}
S(\mathbf{x}) = f_{\text{neural}}(\mathbf{x}) + \lambda \cdot f_{\text{PWM}}(\mathbf{x})
\end{equation}

\subsection{Training Objectives}

\textbf{Contrastive positional learning:} Train with matched compositional pairs:
\begin{equation}
\mathcal{L}_{\text{contrastive}} = -\log \frac{\exp(s(E, E'))}{\exp(s(E, E')) + \exp(s(E, H))}
\end{equation}
where $E, E'$ are compensated sequences and $H$ is scrambled.

\textbf{Position prediction auxiliary task:} Add an auxiliary objective to predict motif positions:
\begin{equation}
\mathcal{L}_{\text{pos}} = -\sum_m \log p(\text{position}_m | \mathbf{x})
\end{equation}

\subsection{Evaluation Extensions}

\textbf{Eukaryotic benchmarks:} Extend MIT to eukaryotic promoters (TATA box, Inr, DPE) and enhancers (TF binding site grammar).

\textbf{Gradient-based attribution:} Use integrated gradients or attention analysis to understand what sequence features models attend to.

\textbf{Fine-tuning studies:} Test whether fine-tuning on promoter data can induce mechanistic understanding.

\end{document}